\documentclass[aps,prb,twocolumn,groupedaddress,showpacs,10pt]{revtex4-1}
\usepackage{graphicx}
\usepackage{color}
\usepackage{amsmath, amssymb}
\usepackage{natbib}
\usepackage{hyperref}
\usepackage{overpic}

\DeclareMathOperator{\Tr}{Tr}
\let\Im\relax
\DeclareMathOperator{\Im}{Im}

\newcommand{\muOhmcm}{{$\mu$Ohm$\cdot$cm}}
\newcommand{\sdashd}{$s\text{-}d$}

\begin{document}
% \title{Anomalous temperature-dependence of resistivity in vanadium-aluminium alloys}
\title{\textit{Ab initio} study of anomalous temperature dependence of resistivity in V-Al alloys}

\author{G\'abor Csire}
\affiliation{Materials Center Leoben Forschung GmbH, Roseggerstraße 12, 8700 Leoben, Austria}

\author{Oleg E. Peil}
\affiliation{Materials Center Leoben Forschung GmbH, Roseggerstraße 12, 8700 Leoben, Austria}

\date{\today}

\begin{abstract}
     V$_{1-x}$Al$_x$ is a representative example of highly resistive metallic alloys exhibiting a crossover
     to a negative temperature coefficient of resistivity (TCR), known as the Mooij correlation.
     Despite numerous proposals to explain this anomalous behavior,
     none have provided a satisfactory quantitative explanation thus far.
     In this work, we calculate the electrical conductivity using an \textit{ab initio} methodology that 
     combines the Kubo-Greenwood formalism with the coherent potential approximation (CPA).
     The temperature dependence of the conductivity is obtained within a CPA-based model of thermal
     atomic vibrations.
     Using this approach, we observe the crossover to the negative TCR behavior in V$_{1-x}$Al$_x$,
     with the temperature coefficient following the Mooij correlation, which matches experimental
     observations in the intermediate-to-high temperature range.
     Analysis of the results allows us to clearly identify a non-Boltzmann contribution
     responsible for this behavior and describe it as a function of temperature and
     composition.
\end{abstract}

\maketitle

\section{Introduction}

%%% Background
%%% Problem setting

Half a century ago, Mooij\cite{Mooij1973} demonstrated that in many high-resistivity alloys, the temperature coefficient of resistivity (TCR) can become negative as the residual resistivity increases in magnitude. 
Moreover, the TCR at room temperature appears to be inversely correlated with the residual resistivity---a relationship known as the Mooij correlation\cite{Mooij1973, Tsuei1983}. 
Essentially, it seems as if the alloy avoids exceeding a certain maximum resistivity level at high temperatures.
Mooij suggested that the observed decrease in the TCR, also related to resistivity saturation\cite{Fisk1976},
was closely linked to the reduction of the mean free path towards a minimum value as
electron scattering increased with rising temperature.
Saturation is observed in most high-resistivity metals and alloys, both crystalline and amorphous\cite{Gunnarson2003}.
Despite extensive discussions over several decades regarding both the saturation and negative TCR\cite{Cote1978, allen1980theory, Tsuei1983, Gunnarson2003, Werman2017, Ciuchi2018}, a clear quantitative explanation has yet to emerge.

%%% Overview of research on the topic

The tendency for the resistivity of metals to increase with temperature is generally understood through Boltzmann transport theory.
In this framework, scattering events that cause finite resistivity are considered statistically independent, and the theory is only applicable when the mean free path is longer than the Fermi wavelength.
Ioffe, Regel\cite{Ioffe1960} and Mott\cite{Mott1972} have predicted the existence of a maximum metallic resistivity,
corresponding to a minimum semi-classical quasi-particle mean free path equal to the inter-atomic distance.
Many mechanisms have been suggested to cause negative TCR\cite{allen1980theory, Gunnarson2003},
including thermal disorder effects on the density of states\cite{Chen1972, Harris1978},
\sdashd{} model of band structure effects\cite{Brouers1972, Brouers1975}, 
weak localization\cite{Jonson1979, Girvin1980, Imry1980, Abrahams1979, Lee1985, Tsuei1983, Kaiser1987, Jayannavar1988, Park2003, Gantmakher2011},
modification of Boltzmann theory to include band mixing\cite{Chakraborty1979}, 
spin fluctuations\cite{Babic1971,Jullien1974},
the appearance of a phonon-assisted conductivity channel\cite{Werman2017},
atomic short range order\cite{Richter1983}, extended Ziman theory\cite{Ziman1961, Faber1965, Frobose1977, Nagel1977, Rapp1981}, and, recently, a polaronic mechanism\cite{Ciuchi2018}
Simpler explanations, such as the Fermi smearing effect combined with local minimum of the energy resolved conductivity at the Fermi level were also proposed\cite{Zhao1989}.

However, most of the theories have been criticized on various grounds.
For instance, it was argued in \citet{Ciuchi2018} that weak localization---one of the most popular
explanations of the negative TCR---cannot be responsible for the effect, because the latter persists
up to high temperatures of hundreds of Kelvins, at which quantum coherence is unlikely to survive.

%%% Specific problem considered

A perfect example of a system displaying negative TCR is a V$_{1-x}$Al$_x$ binary alloy\cite{Alekseevskii1975, Meikap2004}.
This alloy forms a stable bcc solid solution (up to $x \approx 40$~at.~\%) and exhibits
a crossover from the positive to the negative TCR as a function of
Al concentration~\cite{Alekseevskii1975}.
Initially, this behavior was attributed to the localized spin fluctuations.
However, this alloy also shows a negative linear magnetoresistance at low temperatures\cite{Aryainejad1985},
which was considered as a strong argument in favor of the weak localization as the main mechanism responsible for the peculiar transport properties\cite{Stolecki1987, Biswas2004}.

%%% Highlight of the results. I would realy write more about results here, rather than methods.
In the present paper, we shall investigate the temperature-dependent resistivity using \textit{ab initio} simulations of a realistic model for V$_{1-x}$Al$_x$ alloys.
Our methodology allows for the simultaneous treatment of the \textit{ab initio} electronic structure, alloy disorder,
and thermal vibrations on an equal footing.
This framework combines first-principle methods for calculating electronic structure and properties, addressing
alloy disorder and thermal vibrations through the coherent-potential approximation (CPA)\cite{Faulkner1980} within the Korringa-Kohn-Rostoker (KKR) method\cite{spr-kkr,Csire202x},
and employing the Kubo-Greenwood formula of linear response theory for transport calculations\cite{Butler1985, Weinberger1996, Banhart1998a}.
This integrated framework has already yielded exceptional results for the conductivity of both non-magnetic and magnetic alloys\cite{Kdderitzsch2011}.
By applying these advanced techniques, we aim to identify the mechanisms responsible for the peculiar features in the concentration- and temperature-dependent electronic transport properties, providing a more accurate depiction of temperature-dependent resistivity.

\section{\textit{ab initio} description of electronic transport}

We perform calculations using density functional theory (DFT) within the local density approximation.
Electronic transport is described by the KKR-CPA formulation of the Kubo-Greenwood
formula, as derived by Butler\cite{Butler1985}.
By suppressing the energy, spin, and orbital indices, the final result for
the conductivity can be schematically written as a sum of two terms
\begin{equation}
 \sigma_0^{\mu \nu} =-\frac{1}{\pi \Omega}
\sum_\alpha x_\alpha
\Tr\left\{
    \tilde{J}^{\alpha \mu}
    \widetilde \tau J^{\alpha \nu}
    \widetilde \tau
    \right\},
\end{equation}
\begin{equation}
 \sigma_1^{\mu \nu} = -\frac{1}{\pi \Omega}
    \sum_{\alpha \beta} x_\alpha x_\beta % \!\!\!\!\!\!\!\!
    \Tr \left\{
        \tilde{J}^{\alpha \mu} \left[ (1-\chi \omega)^{-1} \chi \right]
        \tilde{J}^{\beta \nu}
    \right\},
\end{equation}
\begin{equation}
\chi_{L_1 L_2 L_3 L_4} = \sum_{\mathbf{k} \in \text{BZ}}
    \widetilde \tau_{L_1 L_2}(\mathbf{k}) \widetilde \tau_{L_3 L_4}(\mathbf{k}) 
    -\widetilde \tau_{L_1 L_2} \widetilde \tau_{L_3 L_4},
    \label{eq:chi}
\end{equation}
where the trace is taken over orbital and spin indices,
$\Omega$ is the volume of the unit cell,
$J^{\alpha \mu} \equiv \Big[ J^{n \mu} \Big]_{n = \alpha}$ denote matrix elements of
the current density operator for an alloy component of type $\alpha$ on a lattice site $n$,
with $\widetilde J^{\alpha \mu}$ being the matrix elements in the alloy representation and
$\mu$, $\nu$ representing Cartesian coordinates; $\widetilde{\tau}$ is the coherent
scattering path operator, while $x_{\alpha}$ is the concentration of species $\alpha$.
The quantity $\chi$ is a 4-index object that should be treated as a matrix in double
indices $(L_1,L_2)$, $(L_3, L_4)$.
The additional factor $(1-\chi \omega)^{-1}$ accounts for the vertex corrections
representing the scattering-in term within the Boltzmann formalism\cite{Butler1985}.
Its importance was shown for the dilute limit in Ref.~\onlinecite{Lowitzer2011}.
In the clean limit $J^{\alpha \mu}$ and $\widetilde J^{\alpha \mu}$
become identical and, therefore, the $\sigma_0$ term is canceled by
the second term in the definition of $\chi$. 
However, the $\sigma_0$ term plays an important role in concentrated solid-solution alloys,
as will be shown below.

We also include the effect of thermal vibrations,
which is mainly responsible for the temperature dependent part of the resistivity. 
We employ the alloy analogy model (AAM) described in Refs.~\onlinecite{Ebert2011,Ebert2015},
which approximates thermal atomic vibrations as an isotropic mixture of atomic displacements from equilibrium positions.
The mixture of shifted atoms is treated within the CPA.
Although the alloy analogy model ignores inelastic scattering events,
it produces quite reliable results for the resistivity
induced by thermal displacements in the region of intermediate-to-high
temperatures\cite{Ebert2015}. 
The AAM calculations are performed as a post-processing step with
the potential obtained during the self-consistency with unperturbed atomic positions.
The Fermi level is adjusted to preserve the number of electrons. The corresponding Fermi level variations
are approximately linear in temperature and reaching about 26 meV at $T = 1000$~K.

\section{Temperature dependent electronic and transport properties of V$_{1-x}$Al$_x$ alloys}

V$_{1-x}$Al$_x$ can be stabilized as a bcc solid solution\cite{Okamoto2012}
up to $x \approx 40$ at.~\% (lattice constant $\sim 3.05$~A),
with the residual resistivity varying approximately linearly with aluminum concentration
reaching 240 $\mu \Omega$cm at a concentration of 37 at.~\% of aluminum\cite{Alekseevskii1975}.
This system displays saturation behavior -- the high-temperature
resistivity appears to approach 190-200 $\mu \Omega$cm for $x \gtrsim 0.2$
Moreover, the temperature coefficient of resistivity
becomes negative for high aluminum concentrations, consistent with the
Mooij correlation \cite{Alekseevskii1975}.

%%%%%%%%%%%%%%%%%%%%%%%%%%%%%%%%%%%%%%%%%%%%%%%%%%%%%%%%%%%%%%%%%%%%%%%%%%%%%%%%%%%%
% Figure: Theoretical and experimental resistivity as function of T
%
\begin{figure}[htb!]
    \centering
        \begin{overpic}[width=0.9\linewidth]{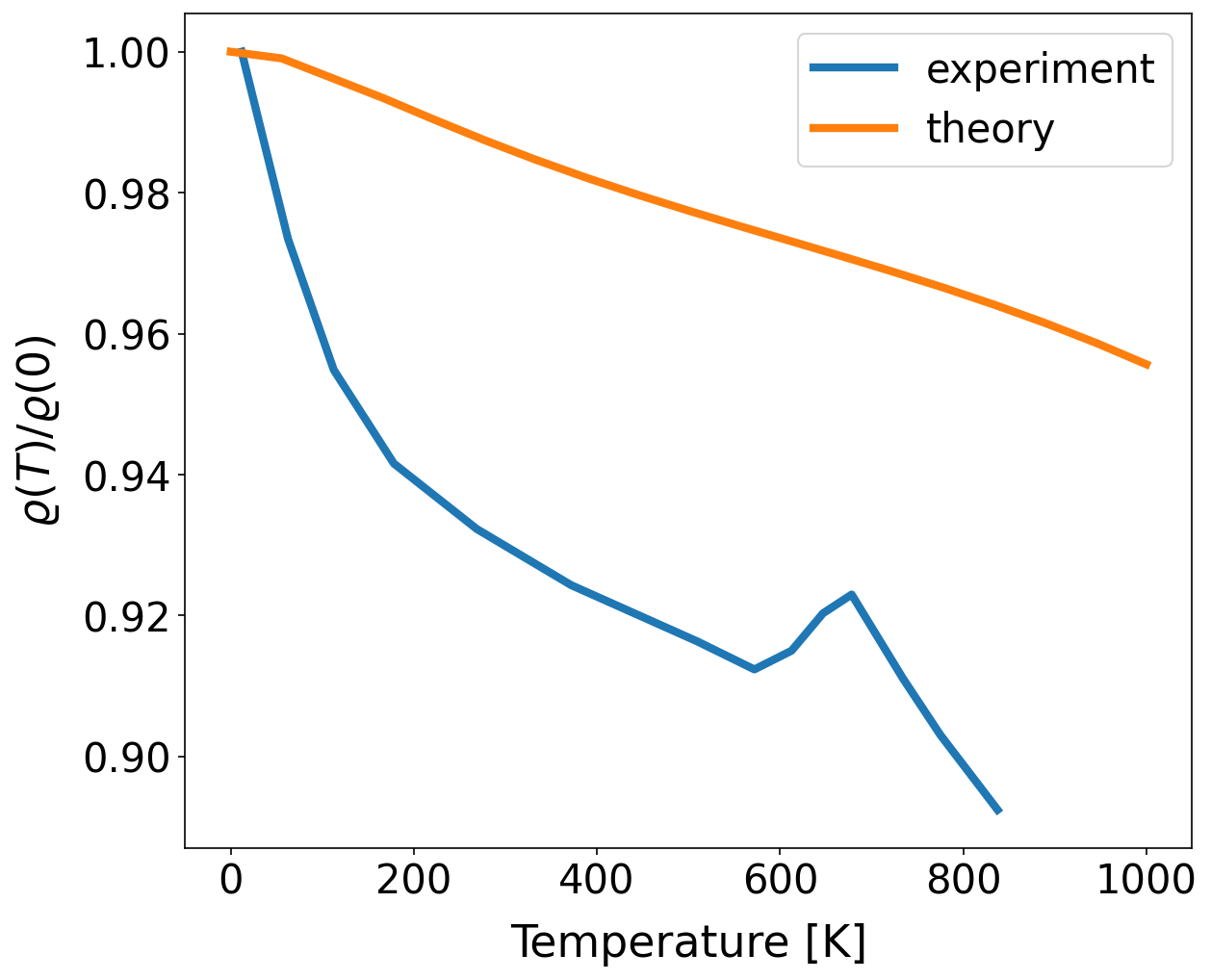}
                \put(0,75){(a)}
         \end{overpic}
         \begin{overpic}[width=0.9\linewidth]{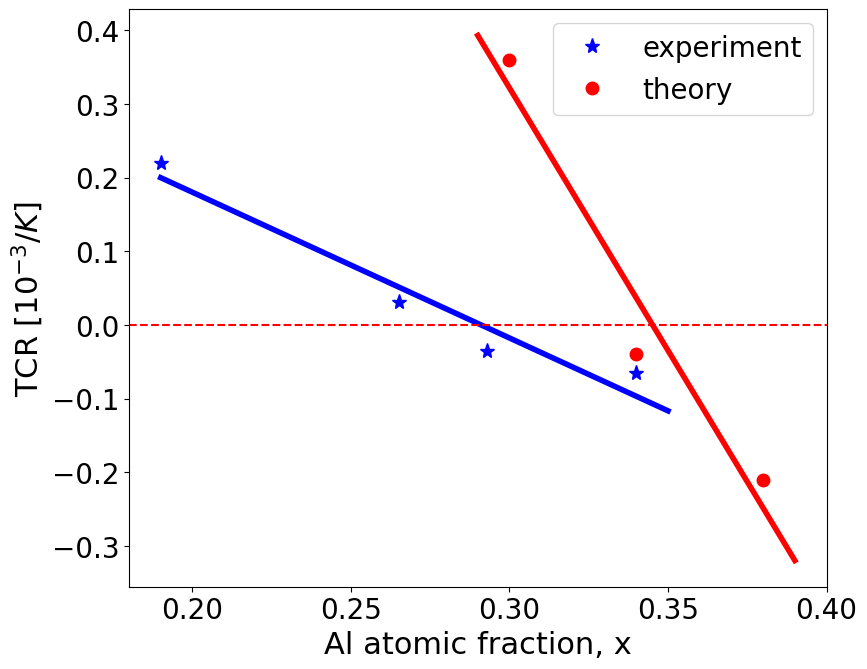}
                \put(0,75){(b)}
         \end{overpic}
    \caption{Comparison of calculated (`theory') and experimental values (a) of the resistivity normalized by the residual resistivity for Al atomic fraction $x=0.34$ and (b) of the temperature coefficient of resistivity (TCR) as a function of Al concentration. Experimental data is taken from Ref.~\onlinecite{Alekseevskii1975}.}
    \label{fig:compare}
\end{figure}
%
%%%%%%%%%%%%%%%%%%%%%%%%%%%%%%%%%%%%%%%%%%%%%%%%%%%%%%%%%%%%%%%%%%%%%%%%%%%%%%%%%%%%

%%%%%%%%%%%%%%%%%%%%%%%%%%%%%%%%%%%%%%%%%%%%%%%%%%%%%%%%%%%%%%%%%%%%%%%%%%%%%%%%%%%%
% Figure: Temperature dependence of \sigma_1, \sigma_0, and total \sigma.
%
\begin{figure*}[htb!]
    \centering
        \begin{overpic}[width=0.95\linewidth]{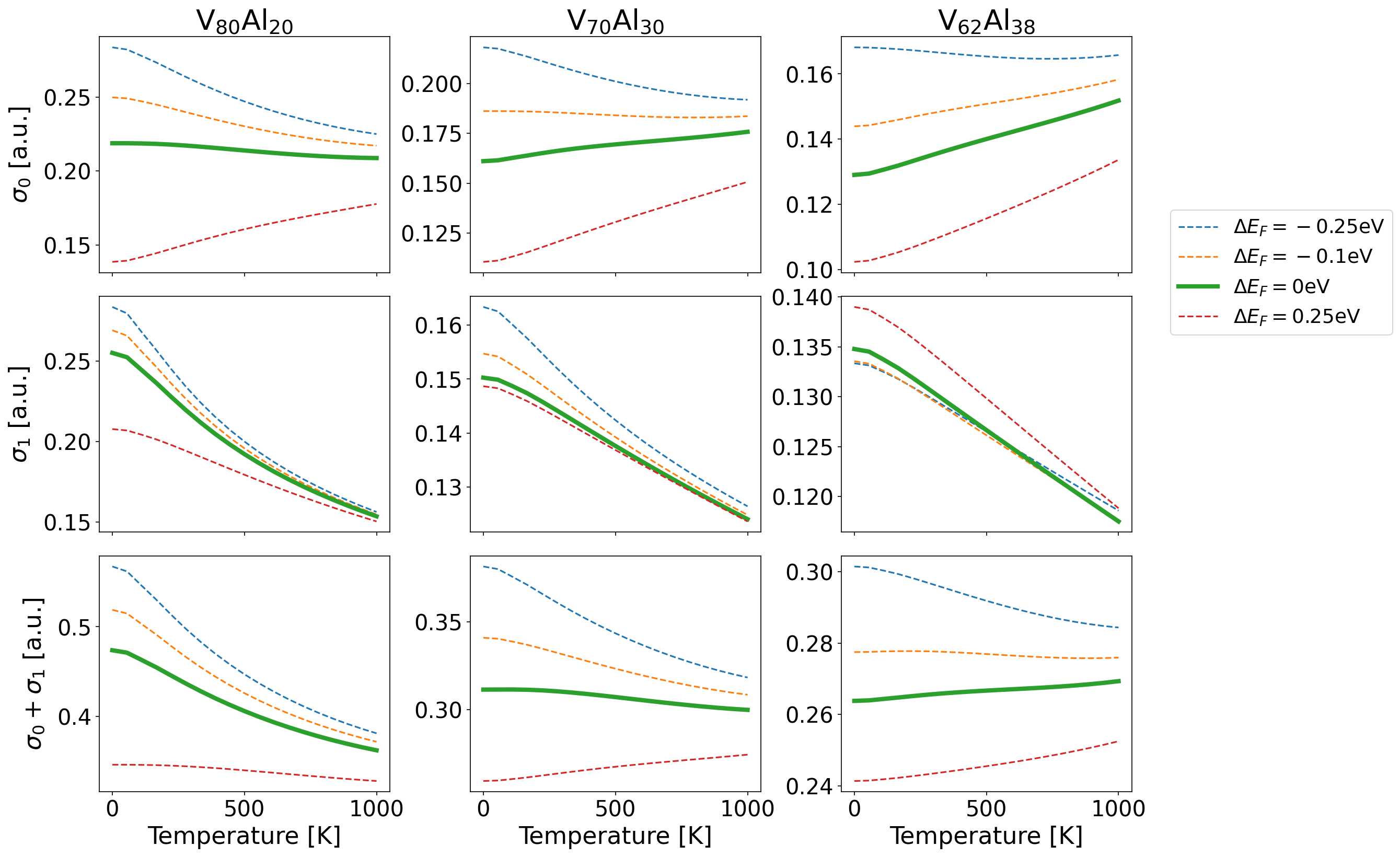}
         \end{overpic}
    \caption{Temperature dependence of the conductivity for V$_{1-x}$Al$_x$  alloys.
        The three columns are related to the Al concentrations $x=0.2, 0.3, 0.38$.
        The first row shows the on-site $\sigma_0$, the second row -- the $\sigma_1$ conductivity term,
        while the third row corresponds to the total conductivity as a function of $T$.
        Several positions of the Fermi levels have been used, with the plots for the true Fermi level
        $\Delta E_F = 0$ shown with thick green lines.
        }
    \label{fig:cond_temp}
\end{figure*}
%
%%%%%%%%%%%%%%%%%%%%%%%%%%%%%%%%%%%%%%%%%%%%%%%%%%%%%%%%%%%%%%%%%%%%%%%%%%%%%%%%%%%%

We have calculated the conductivity of V$_x$Al$_{1-x}$ for several values of $x$
as a function of temperature,
taking into account both thermal atomic vibration and the electronic temperature
(smearing effect of the Fermi-function based on the formula
$\sigma(T) = -\int d\varepsilon \sigma(\varepsilon) 
\partial f(\varepsilon,T)/\partial \varepsilon$).
In Fig.~\ref{fig:compare}(a), we compare the experimental and 
the calculated resistivity (normalized with the residual resistivity) for
$x = 0.34$, where the strongest negative TCR was observed in Ref.~\onlinecite{Alekseevskii1975}.
In experiment, the temperature dependence of the resistivity exhibits three distinct regimes of
behavior at low temperatures (below 200 K), at intermediate temperatures (200-600 K), and
at high temperatures (above 600 K). We see that our calculated temperature dependence
agrees very well with experiment in the intermediate-temperature regime. At higher temperatures,
a phase transition (most likely, precipitation of the intermetallic V$_3$Al phase)
is observed in the experimental sample, resulting in a cusp in the resistivity at around 700 K.
On the other hand, the low-temperature regime is apparently dominated by some other
effects, which are not taken into account by CPA and the model of incoherent atomic vibrations.
We will return to possible other mechanisms later on, in the Discussion section.

In Fig.~\ref{fig:compare}(b), we also compare the calculated TCR to that in experiment. 
One can see that despite the differences in slope (due to an overall underestimation of
the resistivity value in calculations), \textit{ab initio} results show a good
agreement for the crossover concentration (around $x_c \approx 0.3$ in calculations
against $x_c \approx 0.35$ in experiment),
at which the sign change of the TCR occurs, suggesting that
CPA-based calculations capture the fundamental mechanism of the negative TCR correctly.

One of the remarkable results here is that our calculations reproduce quite well
the negative slope of the experimental resistivity in the intermediate temperature range,
despite the lack of any quantum coherence effects, which were often considered as the main
cause of this phenomenon.
To understand the origin of such a behavior in our results, we have analyzed the temperature
dependence of individual terms in the expression for the conductivity.
Specifically, there is a natural separation of the total conductivity into two terms
$\sigma_0$ and $\sigma_1$, which
originates from the configurational averaging, leading to a separate treatment
of on-site and off-site scattering path operators (SPOs)\cite{Butler1985}.
Although $\sigma_1$ also contains an on-site contribution, as can be seen in Eq.~\eqref{eq:chi},
it generally involves statistically independent current matrix-elements for different alloy components,
while $\sigma_0$ is a simple average of terms, each of them related only to a single component.
Details will be discussed in Section~\ref{sec:discussion}, but for now, the crucial point
is that these two terms exhibit very different temperature behavior.

In Fig.~\ref{fig:cond_temp}, we show the temperature dependence of the conductivity contributions
$\sigma_0$, $\sigma_1$ for several concentrations of Al. We have also varied the Fermi level position
to examine its effect on the $T$-dependence.
As one can see, the $\sigma_1$ contribution always decays as the temperature is increased.
This is an expected behavior for the Boltzmann conductivity,
since higher temperature and more intense lattice vibrations enhance scattering,
resulting in the higher resistivity.
Furthermore, $\sigma_1$ is relatively insensitive to the Fermi level.
On the contrary, the behavior of the local term $\sigma_0$ varies significantly,
depending on the composition and the position of the Fermi level.
Importantly, the $\sigma_0$ conductivity term can grow with temperature,
which is especially noticeable at larger concentrations of aluminum or
when the Fermi level is shifted to higher energies.
Moreover, when the value of $\sigma_0$ becomes comparable to that of $\sigma_1$, its anomalous
$T$-dependence can dominate over that of $\sigma_1$, leading to the positive slope of the
total conductivity, as seen in Fig.~\ref{fig:cond_temp} for 38 at.~\% of Al with $\Delta E_{F} = 0$
and for 30 at.~\% of Al with $\Delta E_{F} = 0.25$ eV.
Since doping with Al results in a slight upward shift of the Fermi level,
one can suggest that the latter is the major cause of the anomalous behavior of $\sigma_0$. 

%%%%%%%%%%%%%%%%%%%%%%%%%%%%%%%%%%%%%%%%%%%%%%%%%%%%%%%%%%%%%%%%%%%%%%%%%%%%%%%%%%%%
% Figure: \sigma_0, \sigma_1 as functions of Al concentration.
%
\begin{figure}[htb]
  \centering
    \begin{overpic}[width=\linewidth]{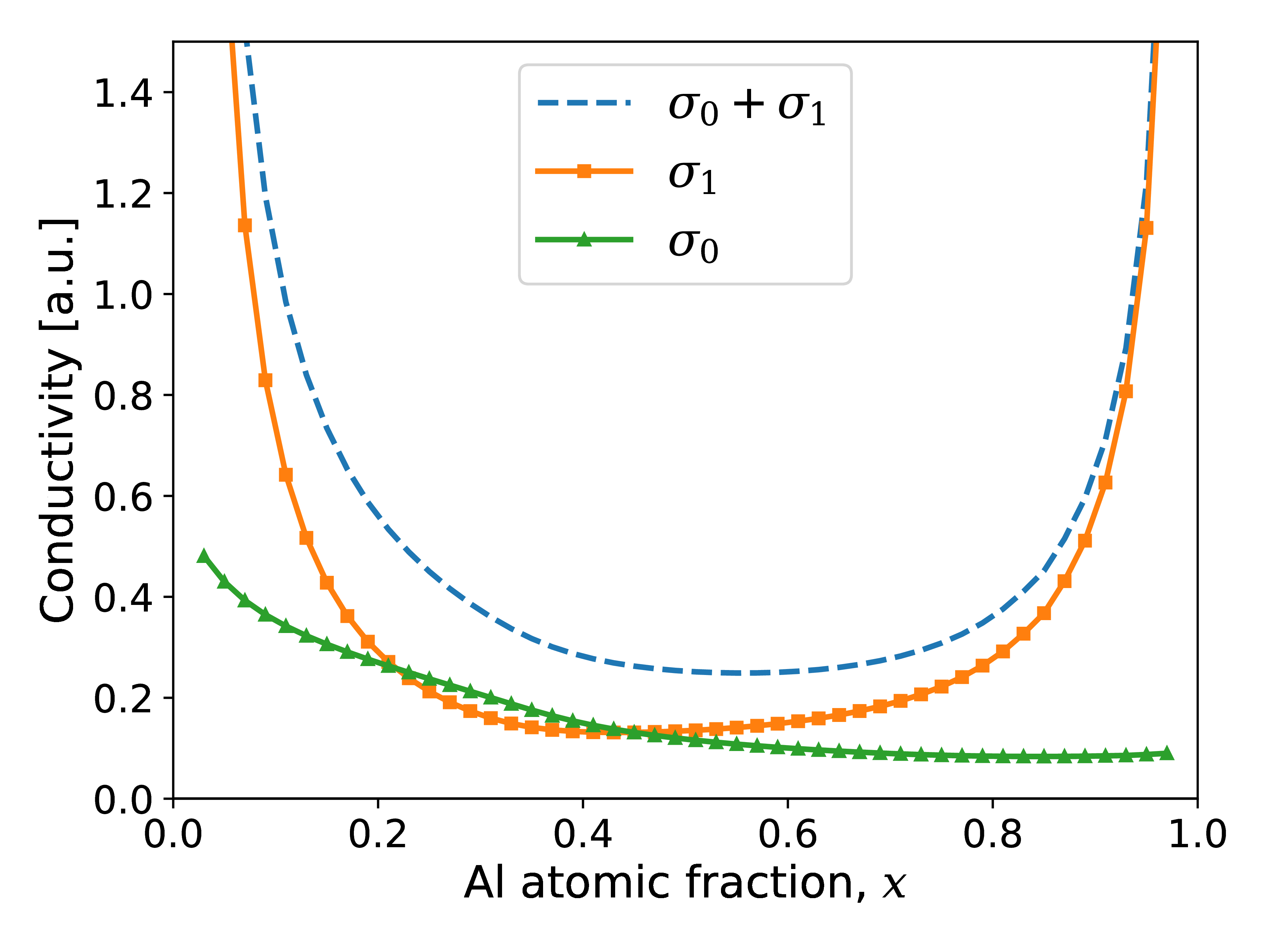}
      % \put(0,75){(a)}
    \end{overpic}
  \caption{Conductivity terms, $\sigma_0$, $\sigma_1$, for V$_{1-x}$Al$_x$ as functions of the Al concentration, $x$, at $T = 0$ K.}
  \label{fig:sigma-vs-conc}
\end{figure}
%
%%%%%%%%%%%%%%%%%%%%%%%%%%%%%%%%%%%%%%%%%%%%%%%%%%%%%%%%%%%%%%%%%%%%%%%%%%%%%%%%%%%%

To investigate the relative magnitudes of $\sigma_0$ and $\sigma_1$, we calculate these two terms
at 0 K for the bcc V$_{1-x}$Al$_x$ solid solution as functions of $x$.
Results are shown in Fig.~\ref{fig:sigma-vs-conc}, where one can see that while $\sigma_1$ shows a typical Nordheim's-law-like shape, the behavior of $\sigma_0$ is very different.
It is rather large (corresponding to $\sim 85-110$ \muOhmcm{}) at the V end,
monotonously crossing over to a smaller value in the Al-rich region.
Interestingly, around 30 at. \% of Al the value of $\sigma_0$ is practically the same as that of $\sigma_1$.
This suggests that $\sigma_0 \simeq \sigma_1$ is one of the necessary conditions for the TCR
to vanish or become negative.
Given that $\sigma_0$ remains finite even in the clean limit and also that it seems to follow
approximately a rule of mixture in the disordered phase, it can be considered as an intrinsic property of elements in a given structure.

\begin{figure}[htb!]
    \centering
        \begin{overpic}[width=0.9\linewidth]{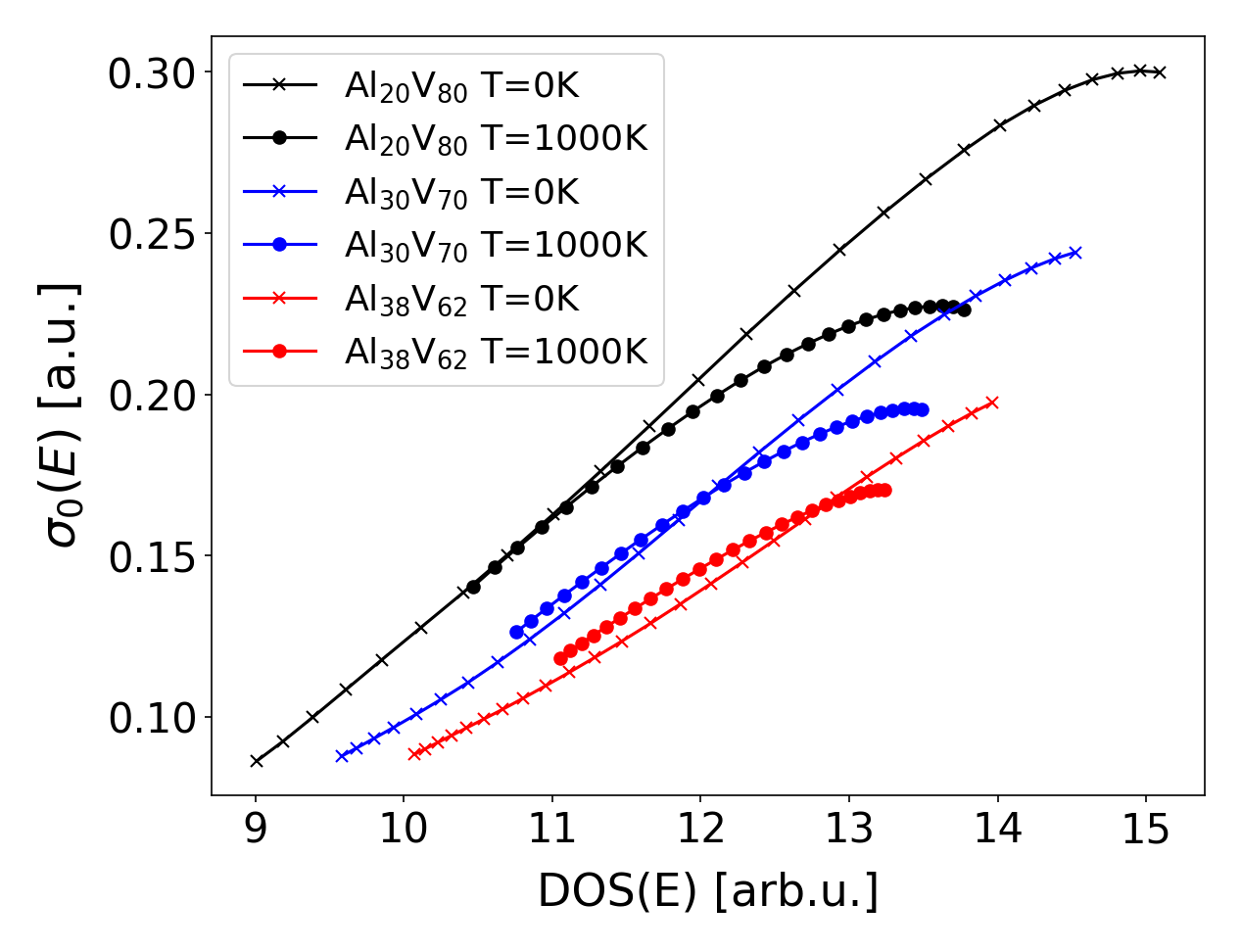}
         \end{overpic}
    \caption{$\sigma_0$ conductivity term plotted against the DOS at the respective
    energy for various V$_{1-x}$Al$_x$ alloys and two temperatures: 0 K and 1000 K.}
    \label{fig:cond_dos}
\end{figure}

The expression for $\sigma_0$, as well as its sensitivity to energy, hints at its possible
relation to the DOS. To check this, we have plotted this term calculated for a range of
energies against the DOS at respective energies.
Results for three compositions and two temperatures are shown in Fig.~\ref{fig:cond_dos},
which reveals a practically linear correlation between $\sigma_0(E)$ and the DOS(E) at 0 K
for a wide range of energies. At high temperatures, the correlation becomes less linear,
but still remains very pronounced. 

%%%%%%%%%%%%%%%%%%%%%%%%%%%%%%%%%%%%%%%%%%%%%%%%%%%%%%%%%%%%%%%%%%%%%%%%%%%%%%%%%%%%
% Figure: \sigma_0 vs DOS
%
%
%%%%%%%%%%%%%%%%%%%%%%%%%%%%%%%%%%%%%%%%%%%%%%%%%%%%%%%%%%%%%%%%%%%%%%%%%%%%%%%%%%%%

%%%%%%%%%%%%%%%%%%%%%%%%%%%%%%%%%%%%%%%%%%%%%%%%%%%%%%%%%%%%%%%%%%%%%%%%%%%%%%%%%%%%
% Figure: Component-resolved DOS for V and Al
%
\begin{figure*}[htb!]
    \centering
        \begin{overpic}[width=0.95\linewidth]{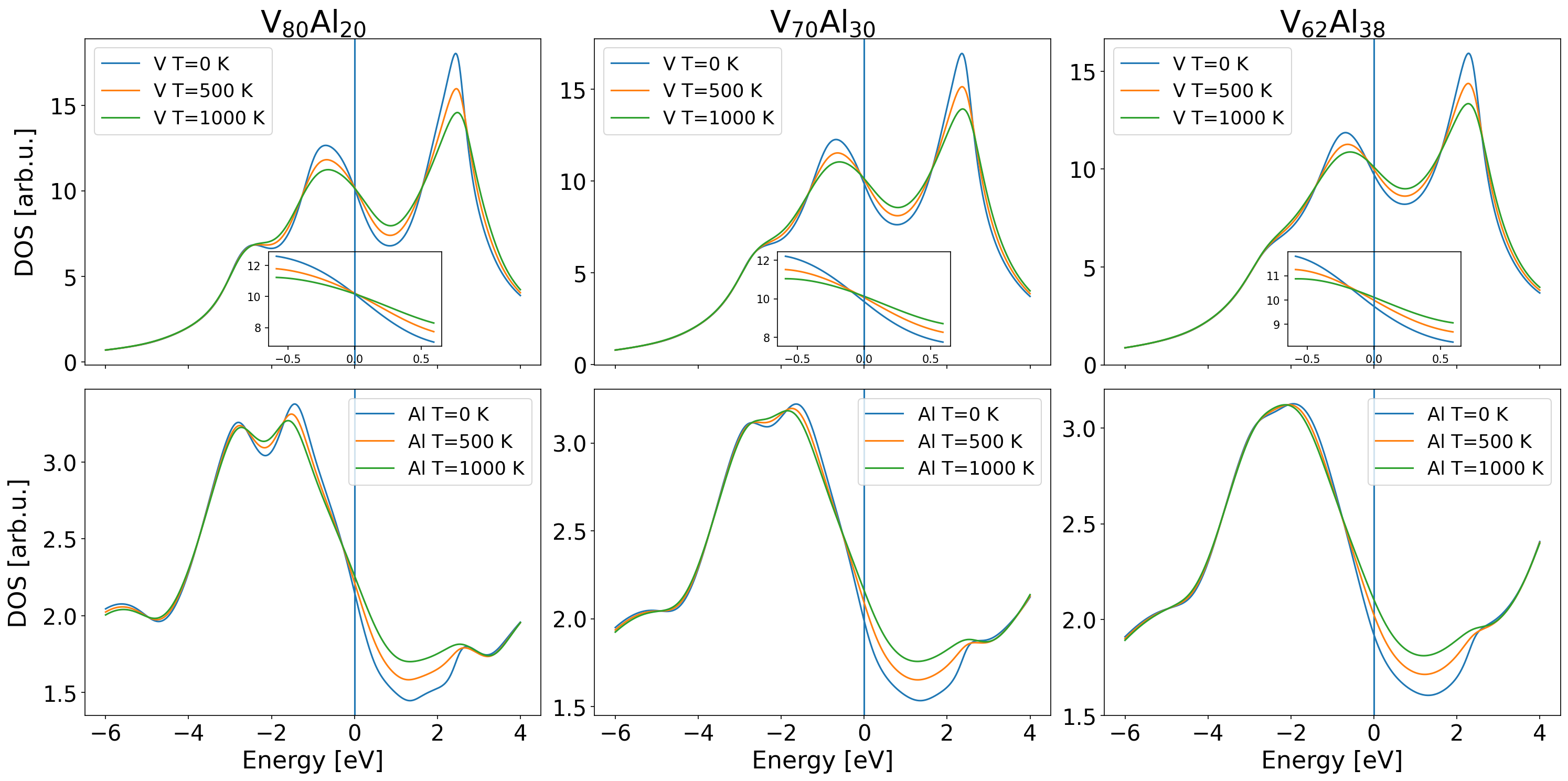}
         \end{overpic}
    \caption{The component-resolved DOS for V$_{1-x}$Al$_x$ alloys
    at three different temperatures 0K, 500K, 1000K.
    The three columns are related to the Al concentrations $x=0.2, 0.3, 0.38$.
    Upper row: The DOS of the vanadium component; inset: Zoom-in around the Fermi level.
    Lower row: The DOS of the aluminum component. }
    \label{fig:dos_temp}
\end{figure*}
%
%%%%%%%%%%%%%%%%%%%%%%%%%%%%%%%%%%%%%%%%%%%%%%%%%%%%%%%%%%%%%%%%%%%%%%%%%%%%%%%%%%%%

The exact origin of this behavior will be discussed later, but already at this stage
we can draw some important conclusions about the temperature dependence of $\sigma_0$.
As one can see in Fig.~\ref{fig:dos_temp}, where individual components' DOS are plotted,
the shape of the partial DOS of both vanadium and aluminum
is only weakly dependent on the concentration of Al.
The main effect of alloying with Al is a slight increase in the occupancy of V $d$ states.
As a result, as the alloy is enriched with Al, the Fermi level, located close to the left peak
in pure V, shifts towards the dip.
Combined with the smearing effect of thermal vibrations, this leads to the tendency for the DOS
at the Fermi level to decrease at low Al content and to grow with increasing temperature,
as one can see, for example, in the insets in Fig.~\ref{fig:dos_temp}.
Given the correlation between the value of DOS and $\sigma_0$,
this observation rationalizes the energy and $T$-dependence of $\sigma_0$ shown
in Fig.~\ref{fig:cond_temp}.
In particular, the origin of the temperature dependence of $\sigma_0$ for $\Delta E_{F} = 0$
can be traced back to the corresponding behavior of the DOS.

%%%%%%%%%%%%%%%%%%%%%%%%%%%%%%%%%%%%%%%%%%%%%%%%%%%%%%%%%%%%%%%%%%%%%%%%%%%%%%%%%%%%
% Figure: DOS, \sigma_0, \sigma_1 as functions of energy at various T
%
\begin{figure*}[htb!]
    \centering
        \begin{overpic}[width=0.9\linewidth]{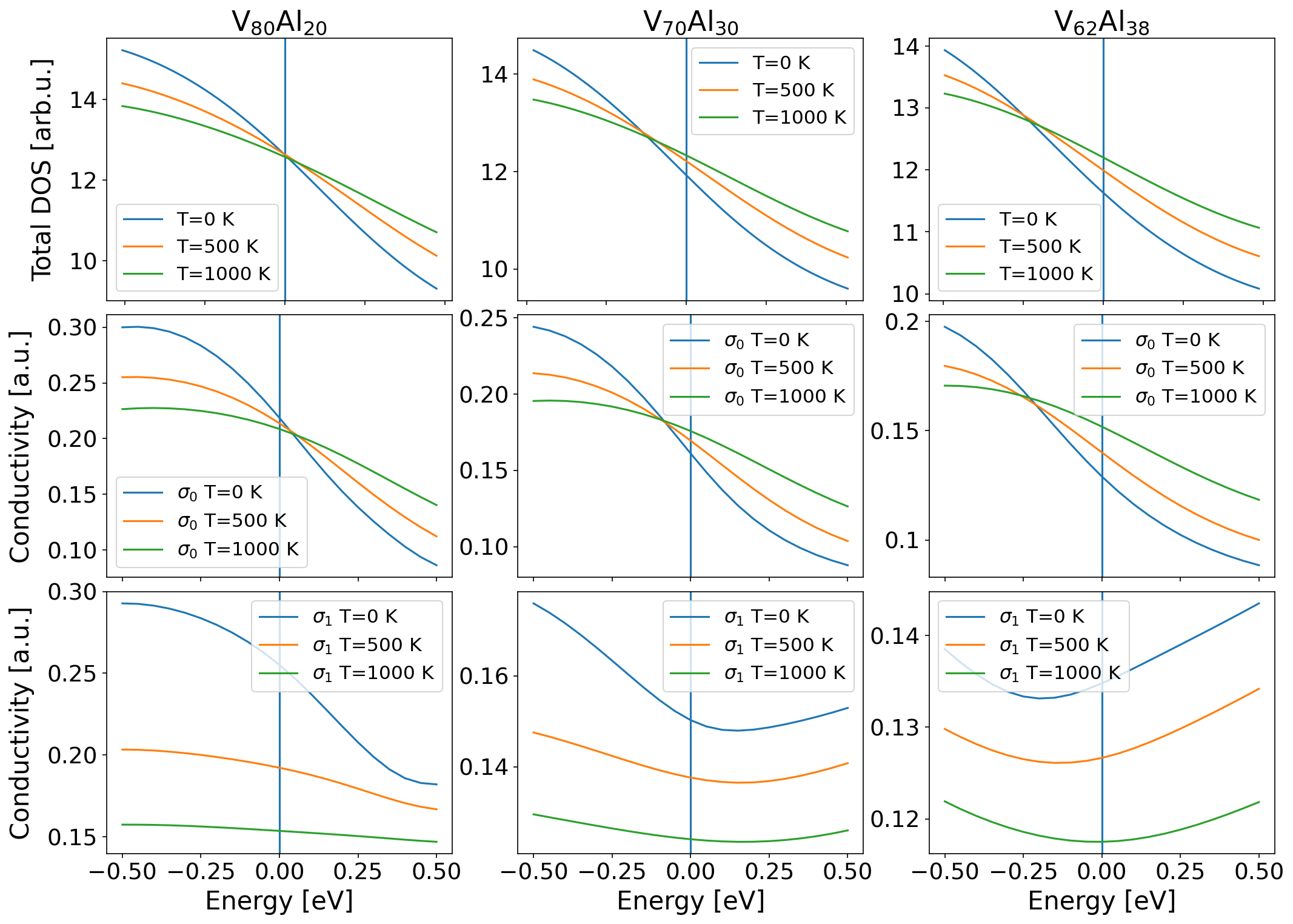}
         \end{overpic}
    \caption{Energy dependence of the DOS and the conductivity for V$_{1-x}$Al$_x$  alloys.
    The three columns correspond to the Al concentrations $x=0.2, 0.3, 0.38$.
    The rows corresponds, respectively, to the DOS, the energy resolved local $\sigma_0$,
    and the energy resolved non-local $\sigma_1$ conductivity term at three temperatures.
    Strong relationship between the DOS and the local $\sigma_0$ conductivity term can be observed.
    }
    \label{fig:cond_energy}
\end{figure*}
%
%%%%%%%%%%%%%%%%%%%%%%%%%%%%%%%%%%%%%%%%%%%%%%%%%%%%%%%%%%%%%%%%%%%%%%%%%%%%%%%%%%%%

To further investigate the relation between $\sigma_0$ and DOS at the Fermi level, we
have calculated $\sigma_0(E)$ for a range of energies around the Fermi level both in low-
and high-temperature regimes.
DOS and the two conductivity terms, $\sigma_0$ and $\sigma_1$, are presented in Fig.~\ref{fig:cond_energy}
as functions of energy. We can see that the DOS and the $\sigma_0$ term are closely related.
At the same time, there is no such relationship between the DOS and $\sigma_1$.
In fact, $\sigma_1$ always decreases with temperature, irrespective of the temperature-induced variations in the DOS.

Furthermore, Fig.~\ref{fig:cond_energy} reveals that the energy resolved $\sigma_1$ has
a local minimum at the energy corresponding to the Fermi level for alloys with Al concentration
between 30 and 38 at. \%.
This feature puts the effect of the Fermi distribution into play which
involves the larger conductivities around the Fermi level with increasing weights
as the temperature is increased.
Here, we mention that the calculation of the resistivity of amorphous metals
(Cu$_x$Zr$_{1-x}$, Mg$_x$Zn$_{1-x}$)
based on real-space Kubo-Greenwood formalism, demonstrated a good agreement with experiment
for the residual resistivity and also revealed a negative T-dependence of the resistivity\cite{Zhao1989}.
The origin of negative TCR was traced back to the fact that $E_F$ is located at or near a local
minimum in the conductivity function. Our calculation also reveals the same condition
for the $\sigma_1$ term in V$_{1-x}$Al$_x$ alloys for the concentration range about
$0.3 < x < 0.4$. Indeed, smearing due to the Fermi-function results in the increase
of $\sigma_1$ with rising temperature,
but this effect is smaller than that of the thermal vibrations and gives rise only to a weak
reduction of the negative temperature trend in $\sigma_1$ when both the Fermi smearing and
the thermal vibrations are taken into account.

\begin{figure}[htb!]
    \centering
        \begin{overpic}[width=0.9\linewidth]{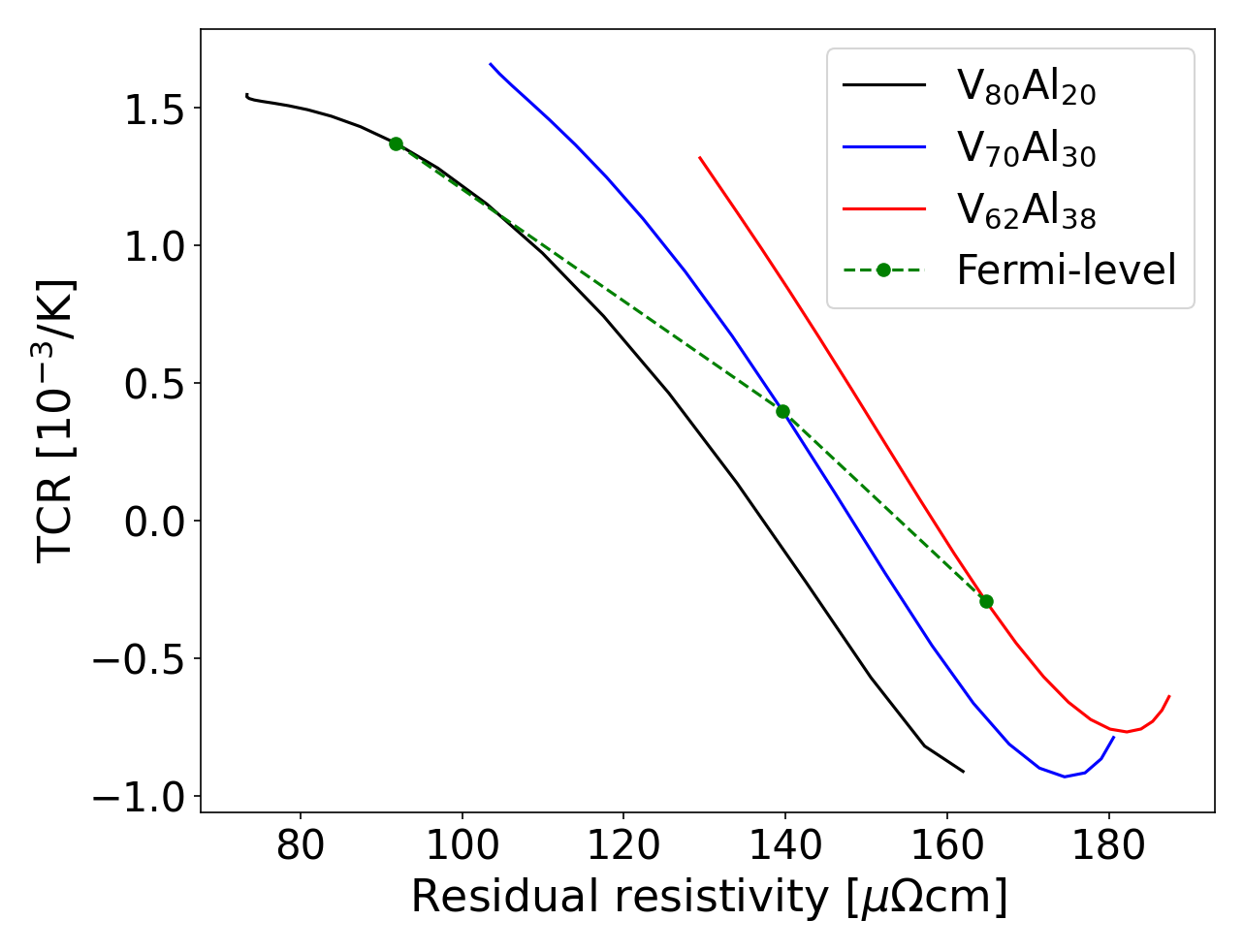}
         \end{overpic}
    \caption{Mooij correlations in V$_{1-x}$Al$_x$ alloys.
    The solid lines are related to the energy resolved resistivity,
    while the dots connected by a dashed line indicate the values at the Fermi level.}
    \label{fig:mooij}
\end{figure}

Another important confirmation of the crucial role of the Fermi level position in determining
the $T$-dependence of the conductivity is a plot of the TCR as a function of the residual
resistivity for various alloy compositions, shown in Fig.~\ref{fig:mooij}.
For each composition, we consider
values of the TCR and the residual resistivity calculated for a range of energies
$[E_F - 0.5, E_F + 0.5]$ eV, with the actual Fermi level marked with dots connected by a dashed line.
Both the resistivity calculated at the true Fermi level and the energy-resolved resistivity
for a fixed composition clearly exhibit the Mooij correlation, which confirms additionally
that the main mechanism is captured correctly by the CPA formalism.

\section{Discussion}
\label{sec:discussion}

The total diagonal conductivity within KKR-CPA can be written as
$\sigma = \sigma_0 - \sigma_{B,0} + \sigma_B$, where $\sigma_{B,0}$ is the local part of the
BZ summation in the Boltzmann term $\sigma_B \equiv \sigma_1 + \sigma_{B,0}$.
The direct relation of $\sigma_B$ to the Boltzmann formalism in the weak scattering limit was
shown in Ref.~\onlinecite{Butler1985}.

To compare the local terms, we write them out explicitly:
\begin{align}
    \sigma_{0}^{\mu\nu} = & -\frac{1}{\pi\Omega} \sum_{\alpha}
        x_{\alpha}
        \Tr \Big[ \widetilde{J}^{\alpha \mu} \widetilde{\tau}
                J^{\alpha \mu} \widetilde{\tau} \Big] = \\
 & =   -\frac{1}{\pi\Omega} \sum_{\alpha} x_{\alpha}
        \Tr \Big[ J^{\alpha \mu} \tau^{\alpha} J^{\alpha \mu} \tau^{\alpha} \Big], \\
%%%
    \sigma_{B,0}^{\mu\nu} = & -\frac{1}{\pi\Omega} \sum_{\alpha \beta}
        x_{\alpha} x_{\beta}
        \Tr \Big[ \widetilde{J}^{\alpha \mu} \widetilde{\tau}
                \widetilde{J}^{\beta \mu} \widetilde{\tau} \Big],
\end{align}
where we have also expressed $\sigma_0$ in terms of bare current operators, $J^{\alpha \mu}$, and
the components' path operators, $\tau^\alpha$, to emphasize that this term is a linear
mixture of component-specific contributions and as such, it is much less sensitive to
chemical disorder than $\sigma_1$. In contrast, the term $\sigma_{B,0}$ cannot
be attributed only to individual components and, furthermore, it scales as $x^2$
with components concentrations and it is expected to vary stronger with increasing disorder.

Such a decomposition allows us to rationalize the general behavior
in the limits of weak and strong scattering.
In the limit of weak scattering, $\tilde{J}^{\alpha\mu} \approx J^{\alpha \mu}$,
and $\sigma_0 \approx \sigma_{B,0}$.
In this case, the total conductivity is dominated by
$\sigma_B$, which is proportional to the scattering time $\tau_B$ and exhibits, thus, a
regular metallic temperature dependence \cite{Butler1985}.

In the limit of strong scattering, the dressed current operator is suppressed, implying
$\sigma_0 > \sigma_{B, 0}$. Moreover, if $\sigma_0$ is sufficiently large and
it increases with temperature, while $\sigma_{B,0}$ decreases, the difference
$\sigma_0 - \sigma_{B, 0}$ will become even larger at higher temperature,
and the overall temperature behavior of the total conductivity can now become anomalous.
The more prominent influence of the $T$-dependence of $\sigma_0$ on the total conductivity
in the strong-scattering regime is also facilitated by the small zero-temperature value of $\tau_B$,
which makes further reductions of the scattering time due to phonon scattering less effective.
This results in the relatively small (negative) temperature coefficient of $\sigma_1$ that can
be easily outweighed by the positive coefficient of $\sigma_0$.

We would like to note in passing that the separation of $\sigma_0$ and $\sigma_1$ terms resembles
the parallel resistor model \cite{Wiesmann1977, Schiller1987, Allen1980, Allen2002},
which was developed to explain the resistivity saturation in A-15 superconductors\cite{Allen1980}.
In this respect, the Kubo-Greenwood formalism within CPA can provide a first-principles
justification for this phenomenological model.

We have, thus far, outlined three key factors necessary for the $\sigma_0$-term to override the
normal Boltzmann temperature dependence of $\sigma_1$:
1) strong scattering, which suppresses $\sigma_{B,0}$;
2) a large value of $\sigma_0$;
3) a positive temperature coefficient of $\sigma_0$.
Since the first point is straightforward, we will focus on points 2 and 3 in the following.

%%%%%%%%%%%%%%%%%%%%%%%%%%%%%%%%%%%%%%%%%%%%%%%%%%%%%%%%%%%%%%%%%%%%%%%%%%%%%%%%%%%%
% Figure: Orbital-resolved \sigma, \sigma_0
%
\begin{figure}[htb]
  \centering
  %% OEP: Made a separate figure.
  % \begin{minipage}[b]{0.425\textwidth}
  %   \begin{overpic}[width=\textwidth]{./VAl_sigma.png}
  %     \put(0,75){(a)}
  %   \end{overpic}
  % \end{minipage}
  % \hspace{0.05\textwidth}
%  \begin{minipage}[b]{0.425\textwidth}
    \begin{overpic}[width=\linewidth]{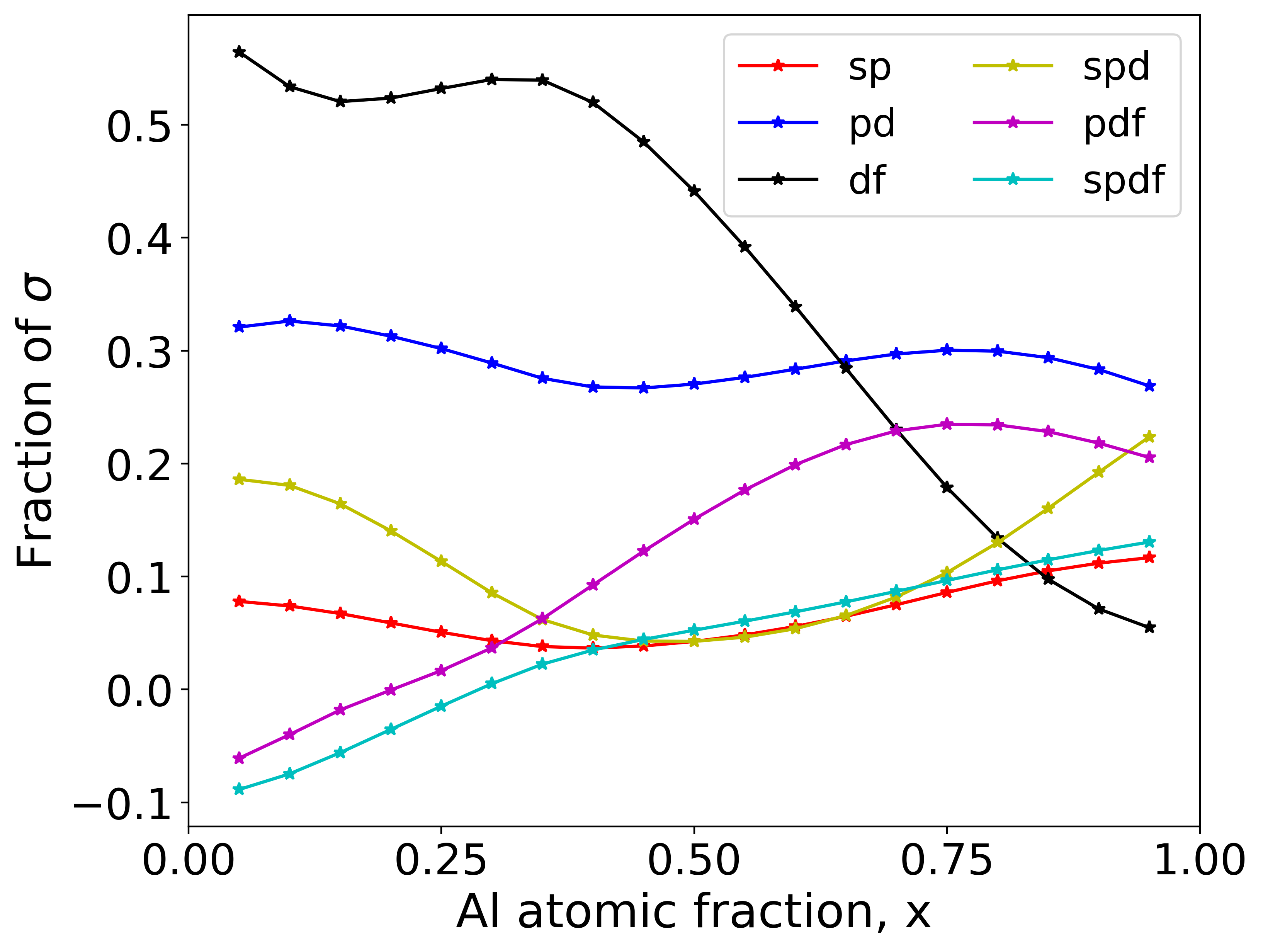}
%      \put(0,75){(b)}
    \end{overpic}
    \begin{overpic}[width=\linewidth]{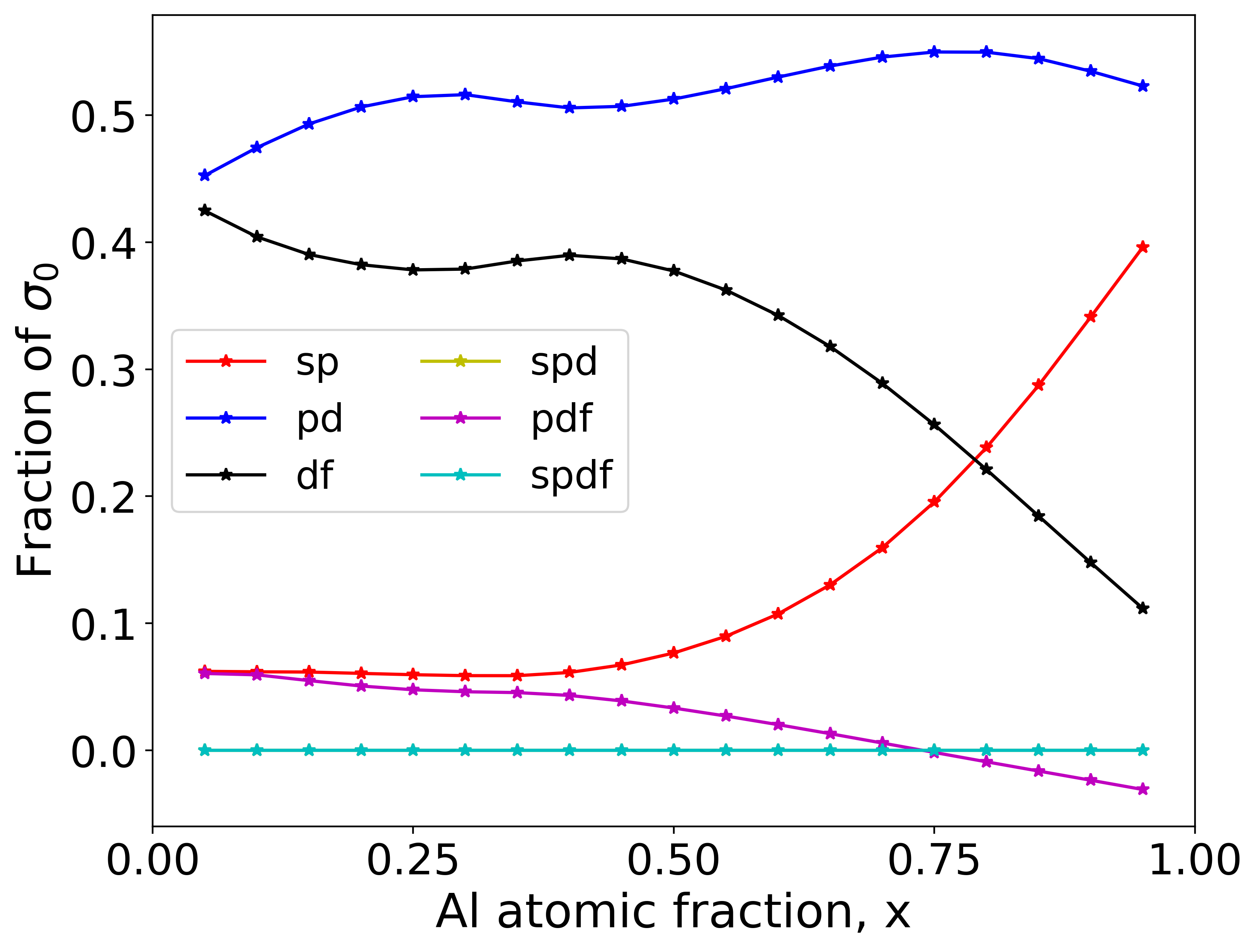}
%      \put(0,75){(b)}
    \end{overpic}
%  \end{minipage}
  \caption{Orbital-resolved contributions normalized by their sum for V$_{1-x}$Al$_x$ as a function of the Al concentration, $x$.}
  \label{fig:sigma}
\end{figure}
%
%%%%%%%%%%%%%%%%%%%%%%%%%%%%%%%%%%%%%%%%%%%%%%%%%%%%%%%%%%%%%%%%%%%%%%%%%%%%%%%%%%%%

As we have shown in the previous section, the temperature dependence of $\sigma_0$ appears to
be determined by that of the DOS at the Fermi level, $E_F$.
In the bcc vanadium-aluminum alloys, the DOS around the Fermi level is associated
predominantly with $d$ electrons, with $E_F$ lying between the peaks of $e_g$
and $t_{2g}$ components. However, only a modest contribution to the current carried by 
these $d$-electron channels is expected based on the Boltzmann picture,
due to the low Fermi velocity and short transport lifetime,
which is confirmed by the high resistivity observed in these alloys.

A better understanding of the nature of conducting channels can be achieved by
considering contributions from various angular momentum channels.
The expressions for the electrical conductivity contain summations over
$L \equiv \{l, m\}$ quantum numbers,
which allows us to define orbital-resolved conductivities by restricting the summation to certain combinations of the angular momentum \cite{Banhart1998}, namely
\begin{equation}
    \sigma_{ll'} \sim \sum_{L_1 L_2 L_3 L_4 \in (ll')} J_{L_1 L_2} \widetilde \tau_{L_2 L_3}  J_{L_3 L_4} \widetilde \tau_{L_4 L_1},
\end{equation}
where the variety of contributions is restricted by selection rules, which
require that $\Delta l$ can only be odd and even for $J$ and $\widetilde \tau$, respectively.
This leaves us with 6 possible combinations for $l_{max} = 3$: $sp, pd, df, spd, pdf, spdf$.

The orbital-resolved contributions have been calculated for $\sigma$ and $\sigma_0$, and the
results are displayed in Fig.~\ref{fig:sigma}.
It is worth mentioning that although thermal vibrations break the local symmetry at finite
temperature, we can still get an insight into the nature of conducting channels by examining
orbital-resolved contributions at zero temperature.

For the total conductivity, three terms, $df$, $pd$, $spd$---all of them involving $d$ states---can
be identified in the top panel of Fig.~\ref{fig:sigma} as the most dominant on the V-rich side.
This can be contrasted with the Al-rich side, where only $p$-orbital contributions are relevant.

At the same time, the largest contributions to the $\sigma_0$ part of the conductivity
(bottom panel of Fig.~\ref{fig:sigma}) are $pd$ and $df$. Moreover, given that $\sigma_0 \sim \sigma_1$
for $0.2 \lesssim x \lesssim 0.5$, one can infer from the figure that the $pd$ and $df$ terms
are mainly due to the $\sigma_0$ contribution to the total conductivity.
This implies that this conducting channel is mainly supported by interorbital
transitions in the vicinity of $V$ atoms, while the Al contribution (e.g., the $sp$ term)
is an order of magnitude smaller.

The selection rules can also explain the nearly linear correlation between $\sigma_0$ and the
DOS at the Fermi level. Indeed, since the path operators are diagonal for $l \leq 2$ and
the current operators only couple orbitals whose $l$-number differs by 1,
the $pd$, $df$ terms are proportional to $\Im g_{d}^V(E_F) \Im g_{l'}^V(E_F)$, with $l' = p, f$,
where $g_l^V$ is the path operator of the vanadium component corresponding to
the angular quantum number $l$ (a contribution from Al is an order
of magnitude smaller, as one can see in Fig.~\ref{fig:sigma}).
Furthermore, $p$ and $f$ of vanadium are mainly determined by the \emph{tails} of states
from neighboring atoms, and their DOS can be considered constant around the Fermi level,
compared with the variations of the $d$-DOS with energy.
Therefore, the variation of $\sigma_0$ will be largely proportional to $\Im g_{d}^V(E_F)$,
explaining the dependence shown in Fig.~\ref{fig:cond_dos}.

The idea that the negative TCR can be traced back to temperature variations of the DOS
at the Fermi level is reminiscent of the interpretation of a mechanism for binary alloys proposed in
\citet{Chen1972}, where a tight-binding model with thermal phonon vibrations
coupled to the on-site potential was considered.
However, it is clear that their picture is incomplete, because it is based on a
single-orbital model and it also predicts that the total conductivity
is proportional to the DOS.
In contrast, in our case, only one
contribution to the total conductivity, namely $\sigma_0$, correlates with the DOS,
and this contribution is intrinsically multi-orbital.

The single-band nature of the model in \citet{Chen1972} was criticized by
Brouers~\textit{et al.}\cite{Brouers1972,Brouers1975}, who
argued that such a strong temperature dependence at the single-particle level can only be
explained by multi-orbital physics and proposed instead a mechanism based on the \sdashd{} model.
This model was originally proposed in Mott's seminal work on the electronic structure
of the Pd-Ag system to explain deviations from Nordheim's law \cite{Mott1935}.
Incidentally, arguments behind the concept of \sdashd{} scattering are somewhat similar
to our argumentation about the proportionality of $\sigma_0$ to the DOS.
The core mechanism of the model is that the incoherent scattering of the $s$-like
conduction electrons into sharp $d$-like states gives rise to an electrical resistivity roughly proportional to the $d$-like DOS at the Fermi level.
Therefore, when the $d$-like DOS drops to a smaller value, the conductivity naturally increases.
However, in our case, a higher $d$-DOS is observed at the Fermi level for increasing temperatures,
which makes us believe that this classical \sdashd{} mechanism is unlikely to 
play a significant role in determining the temperature dependence of the resistivity in VAl. 
Moreover, the \sdashd{} mechanism usually manifests itself in the $\sigma_1$
term, as it was shown for Ag-Pd\cite{Banhart1998} and Au-Pd alloys\cite{Imamnazarov2020}.
In contrast, our results associate the negative TCR with the $\sigma_0$ term,
which is unrelated to the typical elastic scattering described by Boltzmann transport theory.

In our view, a discussion most relevant to our results, was presented in works
of Allen and Chakraborty\cite{Chakraborty1979,Allen1981},
who developed a generalized Boltzmann theory and pointed out that alloys exhibiting
resistivity saturation could be characterized by a significant contribution of interband
transitions to both optical and dc conductivity.
However, multiband generalizations of the Boltzmann transport theory are typically difficult
to handle both analytically and numerically, especially when coupling to thermal atomic vibrations must be taken into account.

Finally, we discuss possible source of errors and deviations from the experimental results,
which leads to underestimation of the residual resistivity.

One potential weakness of CPA is that it does not capture coherence effects and thus,
cannot describe incipient Anderson localization.
However, we do not expect it to play an important role at intermediate and high temperatures
in VAl or similar alloys, because it is a purely
quantum mechanical effect that usually operates at low temperatures.
The more so, supercell calculations of VAl in Ref.~\onlinecite{Brown1989}
did not find any evidence for localized states close to the Fermi level,
which was certified by investigating the participation ratio of eigenstates.

Many transition metals are known to have a substantial Wigner-delay time,
making the treatment of electronic correlations necessary.
In pure vanadium, the electronic correlations can modify the Fermi surface,
as demonstrated in Ref.~\onlinecite{Weber2017}.
This could impact the scattering in the V-rich region of V$_{1-x}$Al$_x$ alloys.
However, it is not likely that it would affect the TCR at high temperatures,
where experimental data shows a linear dependence on temperature.

Spin fluctuations can also be a factor in explaining the negative temperature coefficient of resistivity, as suggested, for instance, in the case of Al-Mn alloys\cite{Babic1971}.
Evidence for spin fluctuations has been found in vanadium from a tunneling study\cite{Gibson1989} and also superconducting density functional theory revealed that taking spin fluctuations into account
is essential to obtain the accurate superconducting transition temperature\cite{Kawamura2020}.
Broadening of energy levels can arise as the consequence of spin fluctuations\cite{Hertz1974},
which can also alter the temperature dependence of the DOS and the resistivity.

We should also mention that in realistic systems, disorder is almost never completely
random. Inhomogeneities may give rise to the appearance of
short- or long-range spatial correlations.
In contrast to long-range order, which always
leads to a reduction of the electrical resistivity, short-range order
can either increase or decrease the resistivity\cite{Tulip2008}. 
For instance, there are a number of alloys, called K-state alloys, where
the residual resistivity decreases if disorder is increased.
In Ref.~\onlinecite{Lowitzer2010}. it was shown that in K-state alloys,
clustering effects increase the $d$-state density at $E_F$, and 
eventually, this leads to an enhanced conductivity.
This suggests that treatment beyond CPA (e.g., with the locally self-consistent Green's function
method \cite{Abrikosov1997,Peil2012}) should be employed to
make more accurate predictions for the resistivity in strong-scattering alloys. 

Another interesting possibility was proposed in Ref.~\onlinecite{Ciuchi2018}, where
a polaronic mechanism of strong disorder renormalization was introduced to
describe how a lattice locally responds to the impurity potential.
This could also be relevant in V$_{1-x}$Al$_x$ alloys,
but such a mechanism definitely requires an \textit{ab initio} approach beyond CPA.

\section{Conclusions}

In summary, we have demonstrated that despite the lack of certain quantum effects,
the Kubo-Greenwood formalism implemented within KKR-CPA is capable of capturing
the negative TCR of highly resistive alloys, and it also reproduces the Mooij correlation
on a good qualitative level. We have identified that the term responsible for this
behavior is the local $\sigma_{0}$ term which does not have an interpretation within
the semiclassical Boltzmann transport theory. Although this term is almost fully
compensated in the weak-scattering limit, this compensation becomes incomplete
in the strong-scattering limit. Unlike the typical Boltzmann conductivity, the temperature
dependence of the $\sigma_{0}$ term is strongly correlated with that of the DOS, which
can be either negative or positive, depending on the position of the Fermi level.
We note that the CPA-based formalism describes the anomalous
temperature dependence of the resistivity in the intermediate-to-high temperature range,
while experiments suggest that additional mechanisms might be at play at low temperatures.
Therefore, other mechanisms proposed earlier are not excluded.

\begin{acknowledgments}
The authors are pleased to acknowledge helpful conversations with
J\'an Min\'ar, Hubert Ebert, Franco Moitzi, Bal\'azs \'Ujfalussy and Andrei V. Ruban.
This research was funded in whole or in part by the Austrian Science Fund (FWF) project
"ReCALL" [10.55776/P33491].
\end{acknowledgments}

\bibliography{refs}
\end{document}